\documentclass{aastex}

\usepackage{emulateapj5}

\shorttitle{Galaxy Magnitudes and Radii}
\shortauthors{Graham et al.}

\begin{document}

\title{Total Galaxy Magnitudes and Effective Radii from Petrosian Magnitudes
and Radii}

\author{Alister W.\ Graham\altaffilmark{1}, Simon P.\ Driver}
\affil{Mount Stromlo and Siding Spring Observatories, Australian National University, Private Bag, Weston Creek PO, ACT 2611, Australia.}
\altaffiltext{1}{Graham@mso.anu.edu.au}

\author{Vah\'e Petrosian}
\affil{Center for Space Science and Astrophysics, Department of Physics, Stanford University, Stanford, CA 94305}

\author{Christopher J.\ Conselice\altaffilmark{2}}
\affil{California Institute of Technology, Mail Code 105-24, Pasadena, CA 91125}
\altaffiltext{2}{NSF Astronomy and Astrophysics Postdoctoral Fellow}

\author{Matthew A.\ Bershady, Steven M.\ Crawford}
\affil{Department of Astronomy, University of Wisconsin, 475 North Charter Street, Madison, WI 53706}

\author{Tomotsugu Goto}
\affil{Institute of Space and Astronautical Science,
  Japan Aerospace Exploration Agency,
  3-1-1 Yoshinodai, Sagamihara, Kanagawa 229-8510, Japan.}
%

\begin{abstract}

%
Petrosian magnitudes were designed to help with the difficult task of
determining a galaxy's total light.  Although these magnitudes (taken
here as the flux within $2R_{\rm P}$, with the inverted Petrosian
index $1/\eta(R_{\rm P})=0.2$) can represent most of an object's flux,
they do of course miss the light outside of the Petrosian aperture
($2R_{\rm P}$).
The size of this flux deficit varies monotonically with the shape of
a galaxy's light--profile, i.e., its concentration.  In the case of a
de Vaucouleurs $R^{1/4}$ profile, the deficit is 0.20 mag; for an
$R^{1/8}$ profile this figure rises to 0.50 mag.  Here we provide a
simple method for recovering total (S\'ersic) magnitudes from
Petrosian magnitudes using only the galaxy concentration
($R_{90}/R_{50}$ or $R_{80}/R_{20}$) within the Petrosian aperture.
The corrections hold to the extent that S\'ersic's model provides a
good description of a galaxy's luminosity profile.
We show how the concentration can also be used to convert Petrosian
radii into effective half--light radii, enabling a robust measure of
the mean effective surface brightness.  Our technique is applied to
the {\sl SDSS} DR2 Petrosian parameters, yielding good agreement with
the total magnitudes, effective radii, and mean effective surface
brightnesses obtained from the NYU--VAGC S\'ersic $R^{1/n}$ fits by
Blanton et al.\ (2005).  Although the corrective procedure described here is
specifically applicable to the {\sl SDSS} DR2 and DR3, it is generally
applicable to all imaging data where any Petrosian index and
concentration can be constructed.

\end{abstract}

\keywords{
galaxies: fundamental parameters --- galaxies: structure
--- methods: analytical --- methods: data analysis 
}

\section{Introduction}

Galaxies are known to possess well--defined correlations between their
various structural and kinematic parameters (e.g., Faber \& Jackson
1976; Tully \& Fisher 1977; Djorgovski \& Davis 1987; Dressler et al.\
1987; Caon, Capaccioli, \& D'Onofrio 1993; 
Graham, Trujillo, \& Caon 2001; De Rijcke et al.\ 2005; 
Matkovi\'c \& Guzm\'an 2005).  These 
empirical ``scaling--laws'' provide key observational constraints
needed to test current theoretical models of galaxy formation and
evolution.  Obviously even the most basic of these, the luminosity--size
relation (e.g., Dutton et al.\ 2005; McIntosh et al.\ 2005), 
relies on our ability to accurately measure robust photometric parameters.

In this vein, the need for a more unified approach to galaxy
photometry has recently been highlighted by Cross et al.\ (2004).
They noted that much of the discrepancy in galaxy magnitudes (and
sizes) between various groups is because of the varying methodology
applied.  For example, some Authors use Kron magnitudes, others Petrosian
magnitudes, some extrapolate fitted models to large radii, while others
use somewhat limited aperture photometry.
This can impact significantly on global measures of the galaxy
population, such as the luminosity function (e.g., Norberg et
al. 2002; Blanton et al.\ 2003a; Driver et al.\ 2005), 
the color--magnitude relation (e.g., 
Scodeggio 2001; Chang 2005, and references therein), and the
luminosity density (e.g., Yasuda et al.\ 2001; Cross et al.\ 2001).
It also, for example, impacts on studies of the supermassive black
hole mass function derived using the galaxy luminosity--black hole mass
relation (e.g., McLure \& Dunlop 2004, Shankar et al.\ 2004, and
references therein).  Perhaps less obvious however, are the
consequences for the calculation of size and surface brightness
distributions (e.g., Kormendy 1977; Cross et al.\ 2001; Shen et al.\
2003; Driver et al.\ 2005).  These are typically derived from the
half--light radius which in turn depends critically on an accurate
assessment of the total flux.  If the magnitude is underestimated,
the size and surface brightness distributions will be affected. 

One of the great strengths of the Petrosian (1976) index, the average 
intensity within some projected radius divided by the intensity at
that radius, and similarly the radii themselves corresponding to some
fixed Petrosian index, is that they do not depend on a galaxy's
distance.  That is, because surface brightness dimming does not change
the shape of a galaxy's light--profile, it does not affect the
Petrosian index, nor does it affect the observed galaxy concentration.
Furthermore, due to the Petrosian index's ability to define
aperture sizes which contain the bulk of an object's light, 
and due to the large influx of small, faint images
of high--redshift galaxies that are now available, the Petrosian index
has experienced a resurgence (e.g., Wirth, Koo, \& Kron 1994; 
Bershady, Lowenthal, \& Koo 1998; Dalcanton 1998; Takamiya 1999;
Bershady et al.\ 2000; Volonteri, Saracco, \& Chincarini 2000; Blanton
et al.\ 2001; Lubin \& Sandage 2001; Conselice, Gallagher, \& Wyse
2002; Yagi et al.\ 2002).
Strauss et al.\ (2002; their section 3.2) do however stress the fact
that a different fraction of galaxy light is missed depending on
whether a galaxy has an $R^{1/4}$ light--profile or an exponential
light--profile, and they emphasize the subsequent need to account for
this in analyses of the Sloan Digital Sky Survey ({\sl SDSS}; York et
al.\ 2000) galaxy data.

In an effort to account for the flux missed by Petrosian apertures,
this paper outlines a corrective procedure to convert Petrosian
magnitudes into total (S\'ersic) galaxy magnitudes, as provided by
popular codes such as GIM2D (Simard 2002), GALFIT (Peng et al.\
2002), and BUDDA (de Souza, Gadotti, \& dos Anjos 2004).  
The key to doing this lies in the `shape' of a galaxy's
stellar distribution, that is, its concentration.  This quantity may
be obtained with or without the use of a fitted light--profile model.


In the absence of measurement errors, all $R^{1/4}$ light--profiles
have exactly the same concentration index $R_{90}/R_{50}$ --- the
ratio of radii containing 90\% and 50\% of the Petrosian flux (e.g.,
Blanton et al.\ 2001; Strauss et al.\ 2002; Goto et al.\ 2003).  The
same is true for an exponential $R^{1/1}$ light--profile, although the
specific value of the concentration will be different in this case.
It follows that the observed range of galaxy concentrations, if not
due to errors, reflects a range of light--profile shapes; that is,
galaxies do not simply have exponential or $R^{1/4}$ light--profiles.

Recognizing this in the {\sl SDSS} data, Blanton et al.\ (2003b) adopted
S\'ersic's (1963, 1968) $R^{1/n}$ model\footnote{A useful compilation of
various S\'ersic expressions can be found in Graham \& Driver (2005).}
to represent the range of galaxy light--profile shapes and provide
estimates of their total luminosities, sizes, and surface
brightnesses.  Indeed, in the case of (dwarf and ordinary) elliptical
galaxies, such an approach is crucial if one is to properly understand
the various relationships between such terms (e.g., Graham \& Guzm\'an 2003,
their Section 4).
However, for low signal--to--noise data or where the spatial
resolution is lacking, it can become difficult to obtain reliable
S\'ersic fits.  One therefore needs an alternative strategy to
obtain the total galaxy flux and associated half--light terms.
 
Using only the observed concentration, $R_{90}/R_{50}$, within the
Petrosian aperture, we provide an easy prescription to recover the
total (S\'ersic) flux from the Petrosian flux while maintaining the
distance--independent qualities of the Petrosian system.  We also
explain how one can recover the effective radii and associated surface
brightness terms.  The corrective formula presented here are not only valid for pure
$R^{1/4}$ or exponential profiles, but applicable to galaxies having
intermediary light--profile shapes and a range of more extreme stellar
distributions.

\section{Petrosian radii and magnitudes} 

The Petrosian (1976) index was initially introduced with the goal of
measuring galaxy evolution.  It gained additional popularity from its
potential to determine the cosmological parameters (e.g., Djorgovski
\& Spinrad 1981; Sandage \& Perelmuter 1990).  Indeed, under the
assumption of structural homology, it provided a means to obtain
``standard rods'' that could be used to constrain cosmogonic models.
Nowadays it is often used as a tool for defining aperture sizes from
which to measure galaxy magnitudes.

The Petrosian index, $\eta(R)$, is a function of a galaxy's
projected radius $R$, and can be written as 
\begin{equation}
\eta(R) = \frac{2 \int_{0}^{R} I(R^{\prime}) R^{\prime} {\rm d}R^{\prime} }{R^2 I(R) } = \frac{L(<R)}{\pi R^2 I(R)} = \frac{\langle I \rangle_R}{I(R)}, \label{EqPet}
\end{equation}
where $I(R)$ is an object's (projected) intensity at some radius $R$, and
$\langle I \rangle_R$ is the average intensity within that
radius.
Following, for example, Blanton et al.\ (2003b), who modeled 183,487 {\sl SDSS} galaxies, 
we adopt S\'ersic's (1963) $R^{1/n}$ model to represent the 
possible range of light--profiles $I(R)$.  This can be written as 
\begin{equation}
I(R)=I_{\rm e}\exp\left\{ -b_n\left[\left( \frac{R}{R_{\rm e}}\right) ^{1/n} -1\right]\right\}, \label{EqSer}
\end{equation}
where $I_{\rm e}$ is the intensity of the light--profile at the
effective radius $R_{\rm e}$, and $n$ defines the `shape' of the
profile.  The term $b_n$ is simply a function of $n$ and
chosen\footnote{The value $b_n$ is such that $\Gamma (2n)=2\gamma
(2n,b_n)$, with $\gamma$ and $\Gamma$ the incomplete and complete
gamma functions respectively (Ciotti 1991).} to ensure the radius
$R_{\rm e}$ encloses half of the profile's total luminosity. 
Using the substitution $x=b_n(R/R_{\rm e})^{1/n}$, and thus
$R = x^n R_{\rm e} / (b_n)^n$ and $dR = (R_{\rm e} n /(b_n)^n)x^{n-1} dx$, 
the Petrosian index reduces to 
\begin{equation}\label{Eqeta}
\eta(x,n) = \frac{2n\gamma (2n,x)}{{\rm e}^{-x}x^{2n}}, 
\label{PetPer}
\end{equation}
where $\gamma $(2n,x) is the incomplete gamma function 
defined as 
\begin{equation}
\gamma (2n,x)=\int ^{x}_{0} {\rm e}^{-t}t^{2n-1}{\rm d}t.
\nonumber 
\end{equation}

\begin{figure}
\epsscale{0.7}
\includegraphics[angle=270,scale=0.7]{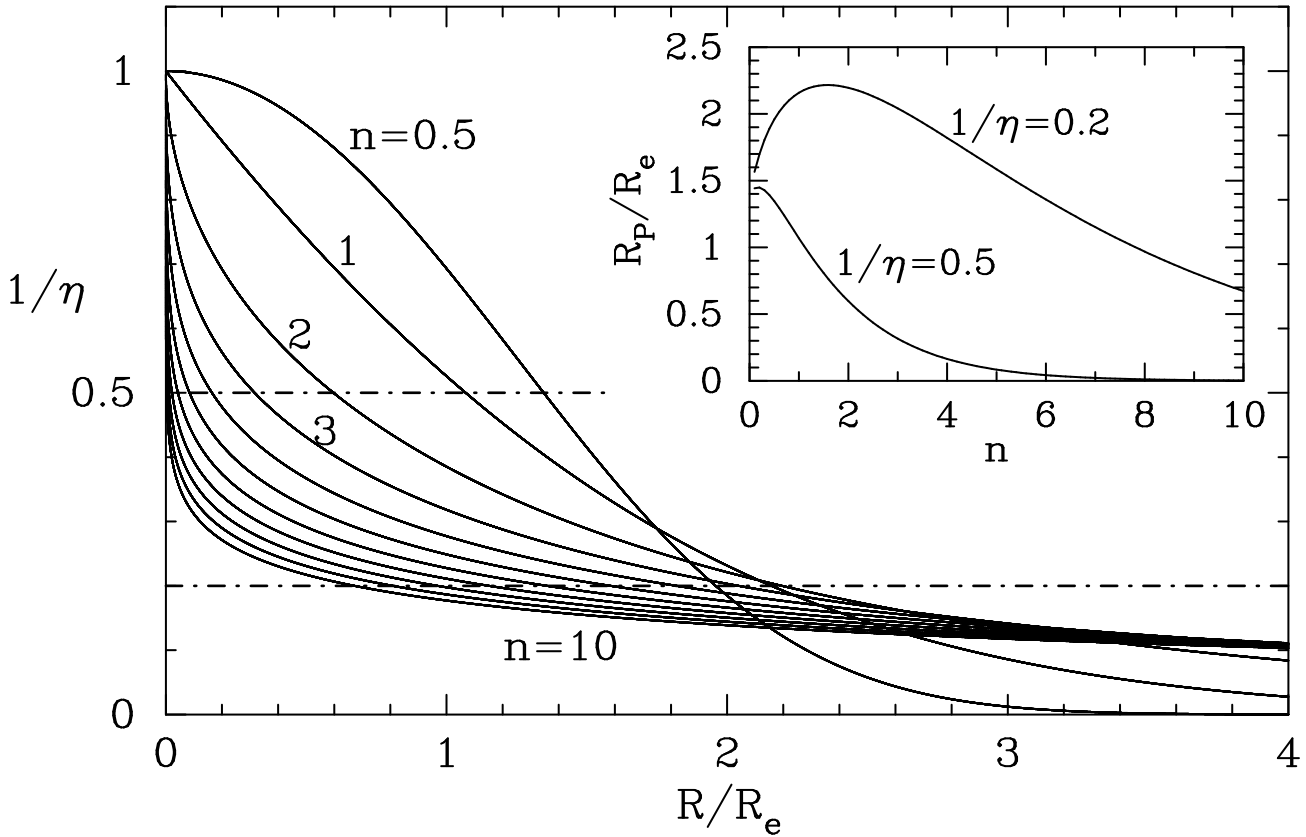}
\caption{
Inverted Petrosian index, $1/\eta$ (equation~\ref{Eqeta}), as a
function of normalized radius, $R/R_{\rm e}$, for light--profiles
having a range of S\'ersic shapes $n$=0.5, 1, 2, 3,... 10 
(equation~\ref{EqSer}).  {\bf Inset Figure:} Number of effective radii
that the Petrosian radius $R_{\rm P}$ --- the radius $R$ at which the
inverted Petrosian index equals some value --- corresponds to when the
$1/\eta = 0.2$ and 0.5 (i.e., the dot--dashed lines in the main
figure).
}
\label{figPet2}
\end{figure}

The inverted Petrosian index, $1/\eta(R)$, is more commonly used in
the literature (e.g., Bershady, Jangren, \& Conselice 2000; 
Blanton et al.\ 2001) 
and is shown in Figure~\ref{figPet2}.  It has a value of 1 at $R=0$
and falls to zero at large radii.  Although the Petrosian index can be
used to determine a galaxy's `Petrosian radius' $R_{\rm P}$ --- the
radius where the index equals some fixed value ---
it is important to determine the relation between this radius 
and the effective radius $R_{\rm e}$: 
as shown inset to Fig. 1, $R_{\rm P}/R{\rm e}$ varies with
S\'ersic index $n$ for a fixed $1/\eta$. 
Constant
multiples of $R_{\rm P}$ have been used as a means to define an
appropriate aperture size for purposes of deriving galaxy magnitudes.
For example, 
some Authors have chosen to use 2$R_{\rm P}$ with $1/\eta(R_{\rm
P})=0.2$ (e.g., Bershady et al.\ 2000; Blanton et al.\ 2001) and
others 3$R_{\rm P}$ with $1/\eta(R_{\rm P})=0.5$ (e.g., Conselice 
et al.\ 2002, 2003).  The {\sl SDSS} consortium adopted the former criteria.
From
Figure~\ref{figPet2}, one can see that different light--profile shapes
reach constant values of $1/\eta$ at different fractions of
their effective half--light radii $R_{\rm e}$.  (The absolute value
of $R_{\rm e}$, and $I_{\rm e}$, are not important, only the ratio
$R/R_{\rm e}$ and the value of $n$ determine $1/\eta$, see
equation~\ref{Eqeta}.)

This is shown more clearly in the inset Figure where one can see, as a
function of profile shape $n$, the number of effective radii that
$1/\eta = 0.2$ and 0.5 correspond to.  If $n$=4, then $1/\eta(R_{\rm
P})=0.2$ occurs at $R_{\rm P} = 1.82R_{\rm e}$.  However, when using
$1/\eta(R_{\rm P})=0.5$, if $n$=4 then $R_{\rm P}$ equals only
$0.16R_{\rm e}$\footnote{Although Section 3.3 of Bershady et al.\
(2000) reports that $1/\eta(R_{\rm P})=0.5$ roughly corresponds to
$R_{\rm P} = 1 R_{\rm e}$, this is only true for S\'ersic profiles
with $n\lesssim 2$ (see Figure~\ref{figPet2}).  Exact values for
$R_{\rm P}$, in terms of $R_{\rm e}$, when $1/\eta(R_{\rm P})=0.5$ are
given in Bershady et al.\ (1998) for n = 0.5, 1, and 4.}.  
Although this particular radius is multiplied by a
factor of three before determining the Petrosian magnitude, this still
amounts to an aperture of size less than 0.5$R_{\rm e}$.  This
particular Petrosian magnitude therefore greatly underestimates an object's
total magnitude.

\begin{figure}
\epsscale{0.5}
\includegraphics[angle=270,scale=0.5]{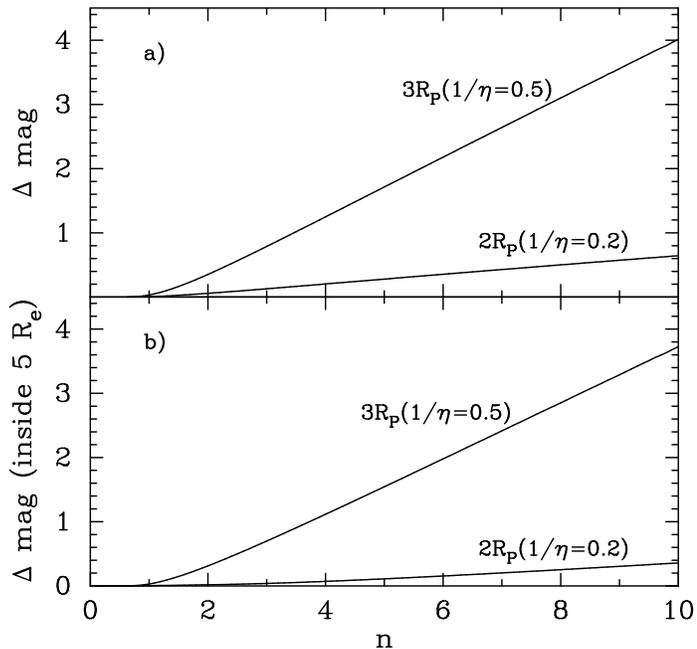}
\caption{
Panel a) shows the difference, as a function of light--profile 
shape `n', between the Petrosian magnitude (equation~\ref{Eqmag}) 
inside (i) twice 
the radius where $1/\eta$=0.2 and (ii) thrice the radius 
where $1/\eta$=0.5 and the total magnitude obtained by 
integrating the $R^{1/n}$ profile to infinity.   Panel b) is 
similar  to panel a) but assumes the $R^{1/n}$ light--profiles 
are truncated at 5$R_{\rm e}$. 
}
\label{figPet4}
\end{figure}

For a S\'ersic profile, the Petrosian magnitude, $m_{\rm P}$, is given
by the expression
\begin{equation}
m(<NR_{\rm P}) = \mu_{\rm e} - 5\log R_{\rm e} -2.5 \log \left[ 2\pi n\frac{{\rm e}^{b_n}}{(b_n)^{2n}}\gamma (2n,x_{\rm P}) \right],
\label{Eqmag}
\end{equation}
where $x_{\rm P}=b_n(NR_{\rm P}/R_{\rm e})^{1/n}$, $N$ is the
multiplicative factor (usually 2 or 3), and $\mu_{\rm
e}=-2.5\log I_{\rm e}$. 
The total magnitude is obtained by replacing $\gamma (2n,x_{\rm P})$
with the (complete) gamma function $\Gamma (2n)$.
%
The extent to which the Petrosian magnitude underestimates the
total magnitude is shown in Figure~\ref{figPet4} under two
conditions.  The first is when the S\'ersic profiles extend to
infinity (panel a).  Although galaxies are recognized not to have
sharp edges, the second condition assumes that the S\'ersic profiles
truncate at 5$R_{\rm e}$ (panel b).  
There is no physical justification for a truncation at 
5$R_{\rm e}$; this is simply chosen in order to explore the 
magnitude deficit if the light--profiles don't extend to infinity. 
While elliptical galaxy light--profiles continue into the background
sky--noise (e.g., Caon et al.\ 1993), there is evidence that some disk
galaxies may truncate at $\sim$4 scale--lengths, or change their
exponential slope at these radii (van der Kruit 2001; Pohlen et al.\
2004; Erwin, Beckman, \& Pohlen 2005), 
but see Narayan \& Jog (2003) and also Bland-Hawthorn et al.\ (2005)
who present a light--profile for the late--type disk galaxy NGC~300 
which extends to 10 scale--lengths. 
%
%
Figure~\ref{figPet4} thus
provides a boundary of sorts to the extent that Petrosian
magnitudes may underestimate a galaxy's total magnitude as a 
function of the underlying profile shape.
To avoid possible confusion, we note that $R_{\rm e}$ shall 
always refer to the value associated with the $R^{1/n}$ model extended
to infinity.  Thus, when the profile is assumed to truncate at 
5$R_{\rm e}$, the value of $R_{\rm e}$ does not change. 

One can see from Figure~\ref{figPet4}a that the use of $1/\eta(R_{\rm
P})$=0.5 results in magnitude differences of 0.5 mag when $n \sim 2.5$;
1.25 mag when $n=4$; 2 mag when $n \sim 5.5$ and considerably worse
for galaxies with yet higher values of $n$.  Obviously such an
approach to determine galaxy magnitudes should be used with caution.
When dealing with dwarf galaxies, because of their faint
central surface brightnesses, the sky flux can often dominate at the
radius where $1/\eta(R_{\rm P})=0.2$.  The use of $1/\eta(R_{\rm
P})=0.5$ is thus more practical, and for galaxies with $n \lesssim 2$ 
the bulk of their flux is still recovered. 
Overall, however, the use of $1/\eta(R_{\rm P})=0.2$ does much better
at recovering a galaxy's true magnitude\footnote{The use of
$1/\eta(R_{\rm P})=0.2$ has also been adopted by some Authors because it
results in a minimal variation of $R_{\rm P}/R_{\rm e}$ with $n$.
For $n \lesssim 4$, this value is around 2, and so to approximate the total 
light, one could measure the light within $R_{\rm P}/2$ and multiply by 2.}.  
When $n$=4, the magnitude
difference\footnote{Bershady et al.\ (2000) 
reported a difference of 0.13 mag, but this is appropriate for an $n=3$ profile
rather than an $n=4$ profile.
The value of 0.13 mag was based on photometry of IRAF
artdata simulations, and therefore is likely to be a round--off or
discretization error in the IRAF rendering.} is only 0.20 mag, 
rising to 0.64 mag when $n=10$.
%
%
Things are better for profiles with smaller values of $n$ and if one
considers the profiles truncate at 5$R_{\rm e}$ (Figure~\ref{figPet4}b).

\section{Recovering effective radii and total (S\'ersic) magnitudes}

In this section we will only consider the case where $1/\eta(R_{\rm
P})=0.2$, using $2R_{\rm P}$ to define the Petrosian magnitude.  The
merits of this precise definition, which was used by the {\sl SDSS} team,
are exalted in Strauss et al.\ (2002, their section 3.2).

\subsection{Concentration}

If one had some way of knowing the underlying light--profile shape
`$n$', then from Figure~\ref{figPet4} one could correct the Petrosian
magnitudes for the missing flux beyond $2R_{\rm P}$.  Given that the
Petrosian index was developed in part to avoid fitting a model to an
object's light--profile, one may prefer not to fit the $R^{1/n}$
model, or indeed one may not have the resolution to do so.
%
Conveniently, central light `concentration' 
is monotonically related to the shape of a light--profile
(Trujillo, Graham, \& Caon 2001a; Graham et al.\ 2001). One
can therefore use concentration as a proxy for the value of $n$.

The {\sl SDSS} consortium have been using a ratio of two radii 
($R_{90}$ and $R_{50}$)
as a measure of an object's concentration, both of which are 
available from the {\sl SDSS} public data releases. 
These radii enclose 90\% and 50\% of the Petrosian flux and their ratio
is shown in Figure~\ref{figPetC} and Table~\ref{Table1} as a function
of the underlying light--profile shape $n$.
%
%
The use of such radii 
avoids the inner seeing--affected part of a light--profile 
and also avoids the outer noisier part of a profile while still
providing a useful range of concentrations for the various galaxies.
For an $n=1$ and $n=4$ profile\footnote{If an $n=1$ and $n=4$ profile
are integrated to infinity, rather than only 2$R_{\rm P}$, then
$R_{90}=2.32R_{\rm e}$ and 5.55$R_{\rm e}$ respectively.},
$R_{90}/R_{50}$ equals 2.29 and 3.42 respectively.  The inverse ratio
is sometimes used, giving 0.44 and 0.29 (Blanton et al.\ 2001, their
section 4.5). 
Not surprisingly, the $R_{90}/R_{50}$ concentration index correlates
with galaxy type and also color (e.g., Strateva et al.\ 2001;
Kauffmann et al.\ 2003). 


\subsection{Magnitudes}

\begin{figure}
\epsscale{0.6}
\includegraphics[angle=270,scale=0.33]{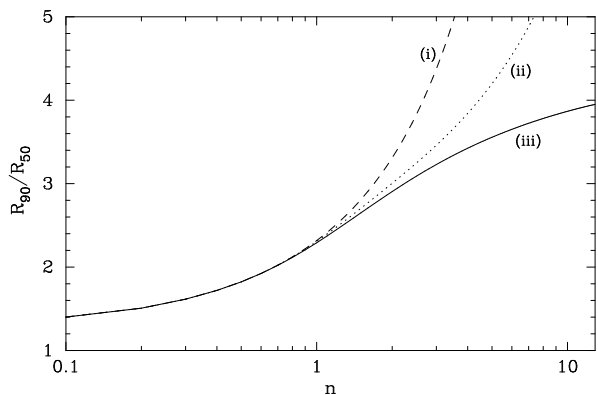}
\caption{
Concentration index, defined as the ratio of radii $R_{90}/R_{50}$, is
shown as a function of the light--profile shape $n$.  (i) $R_{90}$ and
$R_{50}$ are simply the radii containing 90\% and 50\% of the total
flux (obtained by integrating the S\'ersic profile to infinity, note
that $R_{50}=R_{\rm e}$ in this case).  (ii) Similar to (i) except
that the total flux is now considered to be only that within 5$R_{\rm
e}$ (i.e., the S\'ersic profile is truncated at 5$R_{\rm e}$).  (iii)
$R_{90}$ and $R_{50}$ are now the radii containing 90\% and 50\% of
the flux within $NR_{\rm P}$, where we have chosen $N=2$ and $R_{\rm
P}$ is the Petrosian radius such that $1/\eta(R_{\rm P})=0.2$.
}
\label{figPetC}
\end{figure}

Combining Figures~\ref{figPet4} and \ref{figPetC},
Figure~\ref{figPenal}a shows the required magnitude correction in
order to account for the missing flux beyond the Petrosian aperture
2$R_{\rm P}$, when $1/\eta(R_{\rm P})=0.2$, as a function of
concentration within the Petrosian aperture (i.e., curve (iii) from
Figure~\ref{figPetC}).  When $n=4$, $R_{90}/R_{50}=3.42$, and $\Delta
{\rm mag} =0.20$ and 0.07 mag to recover the total flux and that
within 5$R_{\rm e}$, respectively (see Table~\ref{Table1}).

To an accuracy of $\sim$ 0.01 mag, over the S\'ersic interval 
$0.1 < n < 10$, the missing flux (Figure~\ref{figPenal}a) can be
approximated by the expression
\begin{equation}
\Delta m \approx P_1 \exp\left[ \left(R_{90}/R_{50}\right)^{P_2}\right]
\equiv m_{\rm P} - m_x,
\label{EqApprox}
\end{equation}
where $P_1$ and $P_2$ equal 5.1$\times10^{-4}$ and 1.451 in order to
recover the total flux (Figure~\ref{FigApprox}a), and
5.9$\times10^{-5}$ and 1.597 in order to recover the flux within
5$R_{\rm e}$ (Figure~\ref{FigApprox}b).
Here, $m_{\rm P}$ is the Petrosian magnitude and $m_x$ is the 
corrected magnitude which, to an accuracy of $\sim$0.01 mag, is 
equivalent to the S\'ersic magnitude ($m_{\rm S}$).

\subsection{Radii}

The effective radii containing half of the total (S\'ersic) flux
can be computed in a number of ways. 

Figure~\ref{figPenal}b is the combination of Figure~\ref{figPetC} and
Figure~\ref{figPet2}'s inset figure, and allows one to determine the
effective radius $R_{\rm e}$ from the Petrosian radius and
concentration $R_{90}/R_{50}$ within the Petrosian aperture.
Alternatively, from the total galaxy magnitude (approximated by $m_x$) 
one may empirically
determine the aperture containing half a galaxy's light and therefore
obtain the half--light radius this way.  

Alternatively still, for $0.1 < n < 10$, the $R_{50}/R_{\rm e}$ curve
seen in Figure~\ref{figPenal}c can be approximated in terms of
Petrosian concentration by the expression 
\begin{equation}
R_{\rm e} \approx \frac{R_{50}}{\left[ 1-P_3\left( R_{90}/R_{50}\right)^{P_4}\right] } \equiv R_x, 
\label{EqReApp}
\end{equation}
where $P_3$ and $P_4$ equal $8.0\times10^{-6}$ and 8.47,
respectively.
%
%
This approximation is shown in Figure~\ref{FigApprox}c.

\subsection{Surface brightnesses}

\begin{figure}
\epsscale{0.75}
\includegraphics[angle=270,scale=0.75]{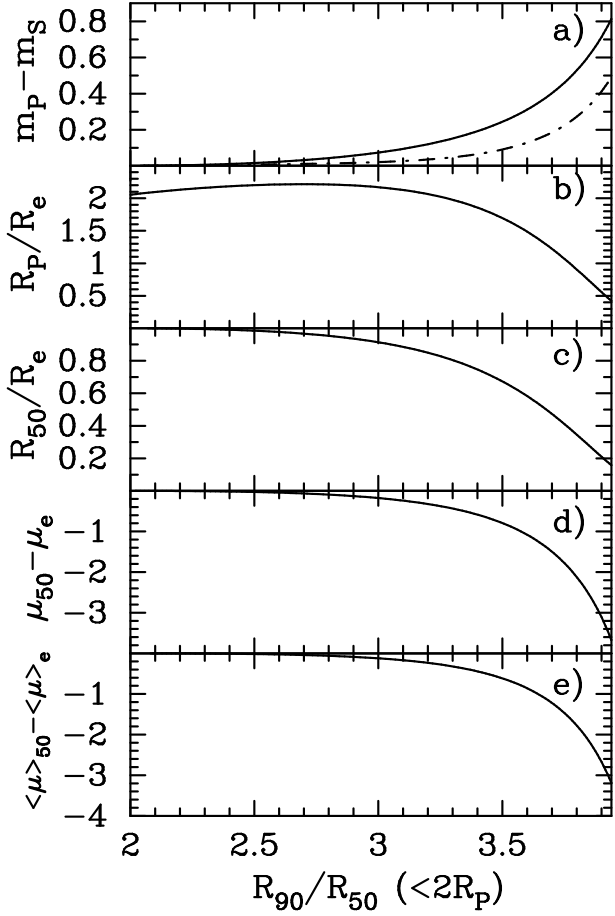}
\caption{
a) Magnitude differences shown in Figure~\ref{figPet4} for the
$1/\eta(R_{\rm P})=0.2$ curves versus the associated Petrosian
concentration index $R_{90}/R_{50}$ within the Petrosian aperture
2$R_{\rm P}$ (as shown by curve (iii) in Figure~\ref{figPetC}).  The
solid curve gives the correction to obtain the total (S\'ersic) magnitude, while
the dash--dot curve gives the correction if the underlying S\'ersic
profile is truncated at 5$R_{\rm e}$.
b) The ratio of Petrosian radii $R_{\rm P}$ (where $1/\eta(R_{\rm
P})=0.2$) to effective radii $R_{\rm e}$. 
c) Radius containing 50\% of the Petrosian flux, $R_{50}$, 
divided by the effective radius $R_{\rm e}$
d) Difference between the surface brightness at $R_{50}$ and
$R_{\rm e}$.
e) Difference between the mean surface brightness within $R_{50}$ and
the mean surface brightness within $R_{\rm e}$.
All panels are relative to $R_{90}/R_{50}$ within 2$R_{\rm P}$ and with
$1/\eta(R_{\rm P})=0.2$ using the ``standard'' definition for the Petrosian 
index (equation~\ref{Eqeta}).
}
\label{figPenal}
\end{figure}

It turns out one can also easily transform $\mu_{50}$, 
the surface brightness at $R_{50}$, 
and $\langle\mu\rangle_{50}$, 
the mean surface brightness within $R_{50}$, 
into the effective surface brightness, $\mu_{\rm e}$, and 
the mean effective surface brightness $\langle\mu\rangle_{\rm e}$. 
From Graham \& Driver (2005, their equation~6 and their section~2.2), one has that
\begin{equation}\label{eqmu}
\mu_{50} - \mu_{\rm e} = \frac{2.5b_n}{\ln(10)}\left[ \left( \frac{R_{50}}{R_{\rm e}} \right)^{1/n} - 1 \right], 
\end{equation}
and
\begin{equation}\label{eqmeanmu}
\langle\mu\rangle_{50} - \langle\mu\rangle_{\rm e} = 
2.5\log\left[ \left( \frac{R_{50}}{R_{\rm e}} \right)^2 \frac{\gamma(2n,x_{\rm e})}{\gamma(2n,x_{\rm 50})} \right], 
\end{equation}
where $x_{50}=b_n(R_{50}/R_{\rm e})^{1/n}$ and 
$x_{\rm e}=b_n$. 
One can see that for a given value of $n$, the ratio
$R_{50}/R_{\rm e}$ (Figure~\ref{figPenal}c) is sufficient to 
solve equations~\ref{eqmu} and \ref{eqmeanmu}. 
Again using the relation between $n$ and $R_{90}/R_{50}$
(Figure~\ref{figPetC}, curve (iii)), Figures~\ref{figPenal}d 
and \ref{figPenal}e
show the surface brightness differences as a function of 
the concentration $R_{90}/R_{50}$. 
In practice, one can directly measure the surface brightness at
$R_{\rm e}$ and/or the mean surface brightness within $R_{\rm e}$, 
the latter of which can also be computed from the expression 
$L_{\rm tot} = 2\pi R_{\rm e}^2\langle\ I \rangle_{\rm e}$. 
Approximations to equations~\ref{eqmu} and \ref{eqmeanmu} 
are therefore not given. 

\vspace{1mm}

Because the above values of $R_{\rm e}, \mu_{\rm e}$, and
$\langle\mu\rangle_{\rm e}$ are those pertaining to a non--truncated
S\'ersic profile, equations~\ref{EqReApp}--\ref{eqmeanmu} are 
independent of any possible profile truncation beyond the Petrosian
aperture.  
%

\section{Application} 

\subsection{Demonstration with {\sl SDSS} data}

The {\sl SDSS} consortium adopted a slightly modified form of the Petrosian
index.  Rather than dividing the average intensity within $R_{\rm P}$
by the intensity at $R_{\rm P}$ (Equation~\ref{EqPet}), they divided
the average intensity within $R_{\rm P}$ by the average intensity
within 0.8--1.25$R_{\rm P}$.  This was done so as to reduce the
sensitivity of the index to noise and (possible) real small--scale
fluctuations in the light--profile (Strauss et al.\ 2002).  As a
result, their definition of the Petrosian index is given by the
expression
\begin{equation}
\eta(x,n) = \frac{\gamma (2n,x_{1.25R}) - \gamma (2n,x_{0.8R}) }
{ \left[ (1.25)^2 - (0.8)^2 \right] \gamma (2n,x) }, 
\label{PetSDSS}
\end{equation}
where $x_{1.25R}=b_n(1.25R/R_{\rm e})^{1/n}$ and 
$x_{0.8R}=b_n(0.8R/R_{\rm e})^{1/n}$. 

\begin{figure}
\epsscale{0.7}
\plotone{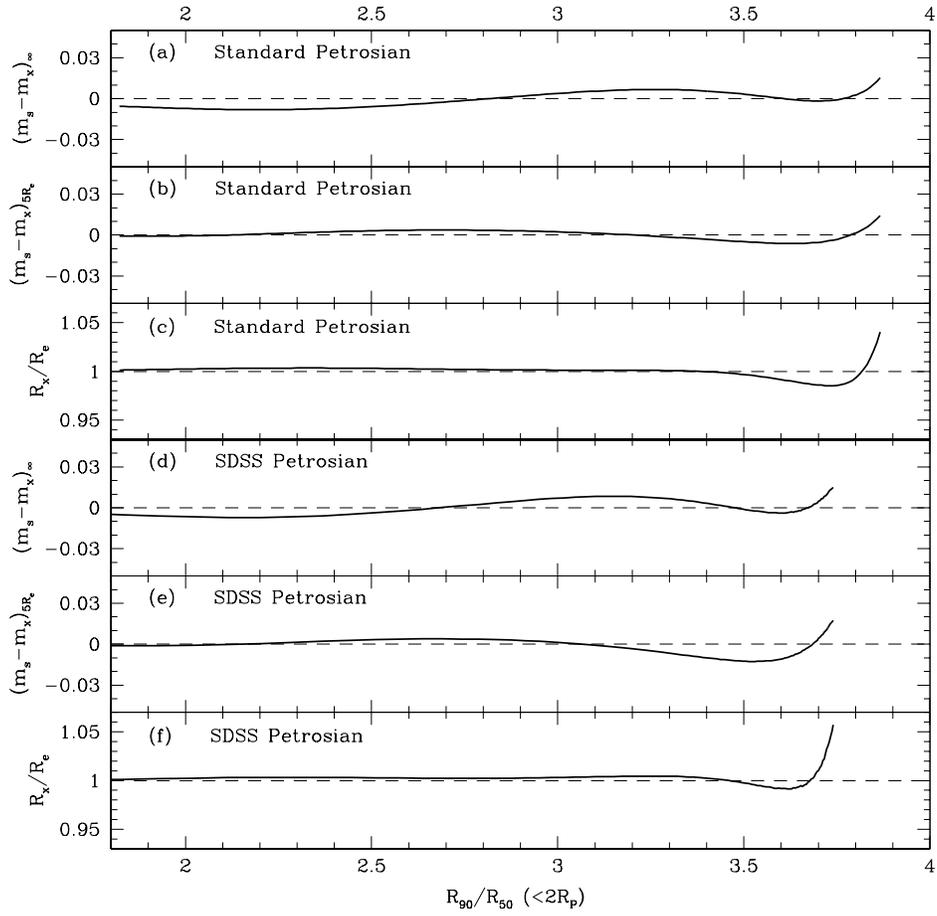}
\caption{
Panel a and b) Difference between the exact magnitude corrections
seen in Figure~\ref{figPenal}a and the approximate 
magnitude corrections given by equation~\ref{EqApprox}. 
%
Panel c) 
Relative error in the radius $R_x$, the 
approximation used to represent the effective radius $R_{\rm e}$
(equation~\ref{EqReApp}). 
Panels d--f)  Similar to panels a--c) but based upon the {\sl SDSS} 
definition of the Petrosian index (equation~\ref{PetSDSS}). 
}
\label{FigApprox}
\end{figure}

In general, where the {\sl SDSS} Petrosian index (Equation~\ref{PetSDSS})
equals 1/0.2, the associated Petrosian radii are slightly smaller than
the radii obtained previously with Equation~\ref{PetPer} equal to
1/0.2 (see Table~\ref{Table1}).  The required magnitude corrections
are therefore slightly greater when using the {\sl SDSS} definition. 
The overall appearance of Figures~\ref{figPet2}--\ref{figPenal} do 
not change, but the exact numbers are different by a few percent and 
therefore provided in the middle section of 
Table~\ref{Table1}.  The approximations given by 
equations~\ref{EqApprox} and \ref{EqReApp} also require a slight modification.
Using the {\sl SDSS} definition of the Petrosian index 
(equation~\ref{PetSDSS}), to recover the total magnitude one now has
$P_1=4.2\times10^{-4}$ and $P_2=1.514$ 
(Figure~\ref{FigApprox}d), and to recover the 
flux within $5R_{\rm e}$ 
$P_1=8.0\times10^{-5}$ and $P_2=1.619$ (Figure~\ref{FigApprox}e).
The values for $P_3$ and $P_4$ are $6.0\times10^{-6}$ and 8.92, 
respectively (see Figure~\ref{FigApprox}f).

We have tested the applicability and need for these corrections using
real data from the {\sl SDSS} consortium.  Specifically, we have taken
the Petrosian magnitudes, 50--percent Petrosian radii, and surface
brightnesses ($m_{\rm P}, R_{50}, \langle\mu\rangle_{50}$) from the
{\sl SDSS} Second Data Release\footnote{{\sl SDSS}--DR2:
\url{http://www.sdss.org/dr2/}} (Abazajian et al.\ 2004) and the S\'ersic--derived total
magnitudes, effective radii, and surface brightnesses ($m_{\rm S},
R_{\rm e}, \langle\mu\rangle_{\rm e}$) from the New York University
Value--Added Galaxy Catalog (NYU--VAGC, Blanton et al.\
2005)\footnote{NYU--VAGC: \url{http://sdss.physics.nyu.edu/vagc/}}.
From this sample we then removed those galaxies whose
Petrosian--derived $R_{50}$ values are biased high by seeing; we
crudely did this by excluding galaxies with $R_{\rm e} <
5^{\prime\prime}$ (90\% of {\sl SDSS} DR2 seeing was between
1.0--1.6$^{\prime\prime}$).
%
In deriving S\'ersic parameters, Blanton et al.\ (2005) prevented the
fits from obtaining S\'ersic indices greater than 5.9 (and less
than 0.2).  To avoid any objects piled up at this boundary, and which 
therefore may not have accurate S\'ersic quantities, we only use
galaxies with $n<5.8$ (and greater than 0.21).  This left us with a
sample of 16128 
galaxies, more than enough to test our method\footnote{For those who need
or wish to use the full galaxy sample, the seeing--affected radii
$R_{50}$ will first need to be corrected (see Trujillo et al.\ 2001b and 
2001c for a prescription to do this).  The S\'ersic radii $R_{\rm e}$
were obtained by Blanton et al.\ (2005) 
from the application of seeing--convolved S\'ersic 
profiles and therefore need no further seeing--correction.}. 

\begin{figure}
\epsscale{0.7}
\plotone{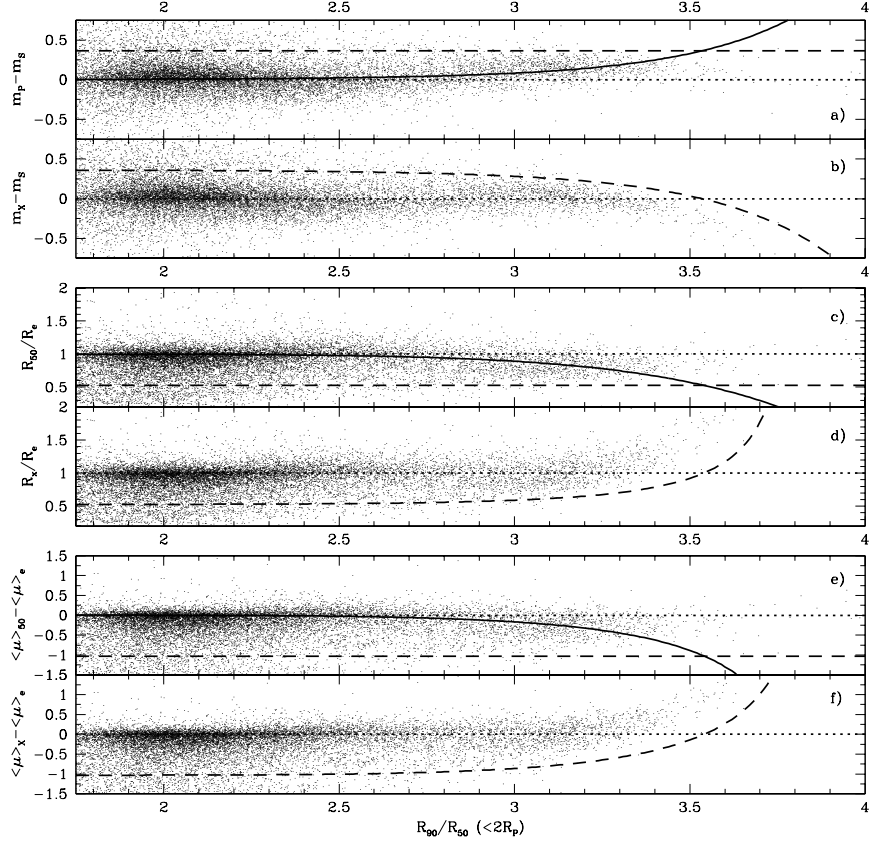}
\caption{
Panel a) The points show the difference between the {\sl SDSS} Petrosian
magnitude $m_{\rm P}$ and the NYU--VAGC S\'ersic magnitudes $m_{\rm
S}$ as a function of concentration for 16128 galaxies.  The solid curve
is the expected difference based on the correction we have formulated
to obtain the flux beyond the Petrosian aperture.  The dashed line 
reflects 
the selection boundary imposed on the data by the restriction that
$n\lesssim5.9$ in the S\'ersic fitting process used by Blanton et al.\ (2003b).
We made a cut at $n=5.8$ to avoid any galaxies which may have piled up at
the upper boundary to $n$. 
Panel b) Our corrected Petrosian magnitudes $m_x$ versus the S\'ersic
magnitudes.  The selection boundary in panel a) has been propagated
into this panel. 
Panel c) Petrosian 50--percent radii $R_{50}$ divided by the S\'ersic
effective radii $R_{\rm e}$ vs. concentration.  From
Table~\ref{Table1}, one can see that the restriction $n$ be less than
5.8 (artificially) prevents $R_{50}/R_{\rm e}$ from getting smaller than
0.53.
Panel d) Our corrected values for $R_{50}$, denoted here by $R_x$, are
shown divided by the S\'ersic radii $R_{\rm e}$.  Again we have
propagated the selection boundary.
Panels e) and f) are similar to the other panels but show the mean 
surface brightness. 
}
\label{FigData}
\end{figure}

%
Figure~\ref{FigData} plots the difference between the Petrosian and
S\'ersic values as a function of the ({\sl SDSS}--tabulated) 
concentration indices $R_{90}/R_{50}$ within the Petrosian apertures.
A similar figure using S\'ersic $n$ rather than concentration is
given in Blanton et al.\ (2003a, their figure 14). 
The solid curves show the expected differences that we have derived.
The dashed curves show the selection boundary imposed upon the data
due to the restriction used by Blanton et al.\ (2005).  Confining $n$
to values smaller than 5.8 can prevent an accurate recovery of the
total S\'ersic flux in large galaxies, and (artificially) prevents
S\'ersic magnitudes getting more than 0.37 mag brighter than the {\sl
SDSS}--defined Petrosian magnitudes (see Table~\ref{Table1}, middle
section).  Of course, measurement errors in either the Pertrosian or
S\'ersic magnitude, and also real deviations from a S\'ersic profile,
can result in differences outside of this selection boundary.  

In Figure~\ref{FigData}a, the S\'ersic magnitudes are clearly brighter
than the Petrosian magnitudes for concentrations greater than about 2.7
($n \sim 1.7$).  In Figure~\ref{FigData}b it is obvious that our
corrected Petrosian magnitudes agree with the S\'ersic
magnitudes, at least to a concentration of $\sim$3.3 ($n \sim 4$) at
which point the selection boundary from the S\'ersic fits start 
to dominate the figure. 
In the concentration bin 
$2.75 \le R_{90}/R_{50} < 3.0$, the mean difference between the Petrosian
and S\'ersic magnitudes, 
$\langle m_{\rm P} - m_{\rm S} \rangle$, 
equals 0.087$\pm$0.005 mag: close to the expected offset of 0.06 mag at 
$R_{90}/R_{50} = 2.87$ (the mean concentration inside this bin). 
In this same bin, $\langle m_{\rm x} - m_{\rm S} \rangle = 0.027\pm 0.005$ mag.
In the next bin, $3.0 \le R_{90}/R_{50} < 3.25$,
$\langle m_{\rm P} - m_{\rm S} \rangle =0.144\pm 0.004$ mag, with the 
expected value at $R_{90}/R_{50} = 3.12$ (the mean concentration inside this bin)
equal to 0.13 mag.  Similarly, 
$\langle m_{\rm x} - m_{\rm S} \rangle= 0.028\pm 0.004$ mag in this bin. 
The mean value of $R_{90}/R_{50}$ in the bin $3.25 \le R_{90}/R_{50} < 3.5$
is 3.34, for which the expected value of $\langle m_{\rm P} - m_{\rm S} \rangle$
is 0.22 mag.  However, the measured value is only 0.148$\pm$0.006 mag.  
The reason for this
is partly due to the selection boundary, but primarily due to the
underestimation of the S\'ersic magnitude by Blanton et al.\ (2005).  
Figure~9 from Blanton et al.\ 
reveals that at $n=4$, corresponding to $R_{90}/R_{50}=3.35$, their S\'ersic
fluxes are underestimated by $\sim$7\%.  This therefore accounts for the 
value of 0.148$\pm$0.006 instead of $\sim$0.22, and for the negative value of 
$\langle m_{\rm x} - m_{\rm S} \rangle= -0.063\pm 0.006$ mag in this bin. 

For a system with $n=5$, Blanton et al.\ (2005) typically obtained
$n=$4.0--4.6 and recovered only $\sim$90\% of the actual flux from
their simulated test galaxies.  The situation is systematically worse for
galaxies with higher values of $n$ and accounts for much of the apparent
over--correction in Figure~\ref{FigData}b at high--concentrations.
Due to this problem, the simple corrective procedure presented here to
allow for the missing flux outside of the Petrosian aperture 
actually provides
a means to acquire (total) S\'ersic magnitudes 
which are more accurate than those obtained from the direct
application of S\'ersic models to the data.
For a typical elliptical galaxy, our method improves the accuracy on 
the galaxy magnitude from a couple of tenths of a mag 
to a couple of hundredths of a mag (assuming that the S\'ersic profile
within the Petrosian aperture continues outside of it). 

We do note that $>$99\% of the {\sl SDSS} Main Galaxy Sample 
($m_r < 17.77$) have $R_{90}/R_{50} < 3.5$ (Blanton et al.\ 2003b; 
Nakamura et al.\ 2003).  Thus, cases where corrections are most 
severe are rare.  
This, however, is not to say that the high--concentration objects are
uninteresting.  Due to the $M_{\rm bh}$--concentration relation
(Graham et al.\ 2001), they are the galaxies expected to have the most
massive supermassive black holes.

In passing we also note that Figures~\ref{FigData}a and \ref{FigData}b
reveal that the {\sl SDSS} Petrosian magnitudes are too faint, in the
sense that they underestimate the total galaxy flux, at the
high--luminosity end.  Correcting for this increases the galaxy number
counts at the bright--end of the (Petrosian--derived) luminosity
function (see Blanton et al.\ 2003a).

Figure~\ref{FigData}c shows the difference between the Petrosian radii
$R_{50}$ and the S\'ersic effective radii.  There is a clear offset at
the high--concentration end of the diagram, albeit constrained within
the selection boundary which prevented S\'ersic indices greater than
5.8 and thus, from Table~\ref{Table1}, $R_{50}/R_{\rm e} < 0.53$.  At
the concentration $R_{90}/R_{50}=3.4$, $n \sim 4.5$ and from Figure~9
in Blanton et al.\ (2005) one can see that their S\'ersic effective
radii are only $\sim$80\% of the true effective radii.  Allowing for
this, the average ratio $R_x/R_{\rm e}$ should be $\sim$1.25 at
$R_{90}/R_{50}=3.4$, exactly as observed in Figure~\ref{FigData}d.
That is to say, the apparent mis--match at high--concentrations is
understood: it arises from the systematics which Blanton et al.\ (2005) 
identified in their fitted S\'ersic quantities. 

Figure~\ref{FigData}e reveals that the early--type
galaxies, defined to be those with concentrations greater than 2.86
(e.g., Shimasaku 2001; Nakamura et al.\ 2003), have Petrosian surface
brightnesses which are clearly brighter than the S\'ersic--derived
surface brightness.  Figure~\ref{FigData}f shows that the 
procedure does a good job at correcting for the missing flux beyond
the Petrosian apertures, at least until a concentration of around 3.3
where the selection boundary and the underestimation of the S\'ersic
flux once more come into play.

\subsection{Application in CAS space}

Another popular setup is that used in the CAS system (Conselice 2003),
which measures the concentration (C), asymmetry (A), and
stellar/star--forming clumpiness (S) of a galaxy's light distribution.
Within the CAS code, Petrosian apertures have sizes of 1.5$R_{\rm P}$,
with $1/\eta(R_{\rm P})=0.2$.  Concentration is also defined slightly
differently from that used by the {\sl SDSS} consortium: it is the ratio of
radii containing 80\% and 20\% (rather than 90\% and 50\%) of the flux
within the Petrosian aperture (Bershady et al.\ 2000).

Theoretically, it doesn't matter what choice of radii one uses to
define concentration.  For example, radii containing 100\% and 50\% of
the flux within the Petrosian aperture would work.  In practice, the
use of a large upper--percentage and a small lower--percentage leads
to a greater range of concentrations and thus a clearer distinction
between different profile types.  However, not using too large an
upper--percentage allows one to work with a less noisy part of the
light--profile.  Not using too small a lower--percentage means that
one is less vulnerable to the effects of seeing.  An analysis of the
optimal concentration index (e.g., Graham et al.\ 2001) is, however,
beyond the intended scope of this paper and will be addressed
elsewhere.  We note that such a search
need not be limited to differing percentages, but could include radii
based on different values of the Petrosian index.  Here we simply
provide the corrections using the popular {\sl SDSS} and CAS definitions for
the concentration index and Petrosian index/aperture.

The various size and flux transformation and corrective terms for the
specific Petrosian setup (1.5$R_{\rm P}$, with $1/\eta(R_{\rm
P})=0.2$) and concentration ($R_{80}/R_{20}$) used by the CAS code of
Conselice et al.\ (2003) are given in the lower section of
Table~\ref{Table1}.  To approximate the total S\'ersic magnitude, over a range
in $n$ from 0.1 to 10, this missing flux $\Delta m$ can be approximated by 
\begin{eqnarray}
\Delta m & \approx & -0.096+0.021(R_{80}/R_{20})+0.0044(R_{80}/R_{20})^2 \nonumber \\
         & \equiv  & m_{\rm P} -  m_x. 
\end{eqnarray}
To obtain the flux within 5$R_{\rm e}$, the following provides 
a good approximation
\begin{equation}
\Delta m \approx -0.012-0.0013(R_{80}/R_{20})+0.00346(R_{80}/R_{20})^2.
\end{equation}
The effective S\'ersic radius $R_{\rm e}$ is given by 
\begin{eqnarray}
R_{50}/R_{\rm e} \approx 0.903 + 0.102(R_{80}/R_{20}) 
        - 0.0276(R_{80}/R_{20})^2  & \nonumber \\
    \hskip-30pt  + 0.00119(R_{80}/R_{20})^3. 
\end{eqnarray}


\subsection{General applicability}

Caon et al.\ (1993) have shown S\'ersic's $R^{1/n}$ model fits
early--type galaxies remarkably well (see their figure~2) down to
faint surface brightness levels ($\mu_B \sim 27$ mag arcsec$^{-2}$).
Spiral galaxies, however, usually have two clearly distinct
components, namely a bulge and a disk.  It is therefore pertinent to
inquire how the above corrections may apply when a galaxy is clearly
better represented by two components, as is the case for the
intermediate--type galaxies.

When dealing with luminous galaxies, the rough divide between
early--type and late--type galaxies has been taken to occur at
$R_{90}/R_{50}=2.86$ $(n\sim 2)$ (Shimasaku 2001; Nakamura et al.\ 2003; Shen et al.\
2003, using the {\sl SDSS} definition of Petrosian index and 
aperture).\footnote{Other Authors have used $n=2.5$ (e.g., Blanton et al.\ 2003b) 
or $R_{90}/R_{50}=2.6$ (e.g., Strateva et al.\ 2001; Kauffman et al.\ 2003) to 
mark the rough divide between luminous early-- and late--type galaxies.}
It should be kept in mind that this concentration based
definition of galaxy type is only applicable to luminous systems.
Dwarf elliptical galaxies have exponential--like profiles and thus low
concentrations (e.g., Graham et al.\ 2001; their figure 8), but are
obviously not late--type galaxies.
%
%
The magnitude correction for galaxies having concentrations of this
size or smaller is less than 0.06 mag (Table~\ref{Table1}).  In
late--type galaxies, the exponential disk, rather than the bulge,
dominates the flux in the outer parts.  Whether or not these disks
continue for many scale--lengths or truncate at a few scale--lengths,
if one applies the corrections presented here, one will not
over--correct the galaxy flux by more than 0.06 mag for the
2--component late--type galaxies, nor overestimate the half--light
radius by more than 7\% (Table~\ref{Table1}).
%
%
If, however, one is able to identify the disk--dominated galaxies from
their blue restframe colors, 
or indeed from their low concentrations 
(assuming no low luminosity elliptical galaxies reside in one's
sample, or folding in asymmetry to separate the galaxy classes), 
then one may instead elect to apply the small corrections
applicable for systems with an $n$=1 (exponential disk) light--profile.  Given the
corrections for such stellar distributions are small, a simple answer
may then be to only apply the various corrections to the red galaxy
population, or to systems with ({\sl SDSS}) concentrations greater than say
2.86.

We caution that intermediate objects, such as lenticular galaxies, 
with half their light coming from 
their disk and half from their bulge may not be well approximated with
a single S\'ersic profile.  In such cases the concentration
prescription given here should be used with care as it is intended for
systems which can be approximated with a single S\'ersic function.
In general, however, Section 4.1 and in particular Figure~\ref{FigData} 
reveal that the method developed here seems to work rather well.

A different issue pertains to how one actually measures the Petrosian
index.  For a disk system, the Petrosian index is independent of disk 
inclination if one uses appropriate elliptical apertures reflecting
the inclination of the disk.  If one instead measures the index from
the slope of the major--axis light--profile, such that $ 2/\eta = d
[\ln L(R)]/d[\ln R]$ (Gunn \& Oke 1975), then again the index is not
dependent on the inclination of the disk.
However, the use of circular apertures applied to highly--inclined
disk galaxies will result in erroneously high concentrations compared
to what these values would be if such galaxies were viewed with a
face--on orientation.  Although we have not quantified this effect,
comparisons between eye--ball morphology and concentration index
yields reasonable agreement for the late-type galaxies (e.g.,
Shimasaku et al.\ 2001), suggesting that in practice this effect is
not a significant problem.

Lastly, we note that if image distortion due to seeing is not
corrected for and is such that the PSF has altered the Petrosian 
50\% radius $R_{50}$ and thus $R_{90}/R_{50}$ (or
$R_{80}/R_{20}$) to the extent that the measured value no longer
reflects the intrinsic galaxy concentration, then obviously one should
not be including such objects in plots or analyses that use this
quantity.  The concentration--based corrections outlined in
this study cannot be applied in such circumstances, at least until the
radii are corrected for seeing (e.g., Trujillo et al.\ 2001b,c)
We also direct readers to 
Blanton et al.\ (2001) who present a study of how the {\sl SDSS} 
Petrosian magnitudes are affected by seeing, and Wirth et al.\ (1994) 
who considered the affects of both poor image 
resolution and low signal--to--noise on image concentration. 
%
%
%


\section{Summary} 

In commenting on the adopted Petrosian system of the {\sl SDSS}
consortium, Strauss et al.\ (2002) wrote: ``Any scientific analysis
that uses the [SDSS] redshift survey must consider how the fact that a
different fraction of light is included for elliptical and spiral
galaxies affects the result''.  This was both a recognition that a) their
particular Petrosian system likely captures a different fraction of
galaxy light compared to other surveys and that b) any practical
definition of Petrosian parameters will inevitably miss some fraction
of a galaxy's light, and a different fraction in different galaxies.
For example, a galaxy with an exponential light--profile will have
99\% of its total flux encompassed by the {\sl SDSS} Petrosian
magnitude, but if a galaxy has an $R^{1/4}$ profile, then the flux
beyond the {\sl SDSS} Petrosian aperture accounts for an additional
0.22 mag, i.e., roughly 1/5 th of that galaxy's total light.  In this
paper we outline a simple method for accurately recovering the missed
flux.

Differences between galaxy light--profile shapes result in a range of
galaxy concentrations.  We have shown how these can be used to correct
for the flux outside of the Petrosian apertures and thus provide a
galaxy's total (S\'ersic) magnitude.  The corrections presented here
hold to the extent that S\'ersic's $R^{1/n}$ model provides a good
description of the various galaxy light--profiles.  For early--type
galaxies this is known to be the case (e.g., Caon et al.\ 1993), while
for late--type galaxies the correction is $<$0.06 mag and one can
argue whether or not to apply it.

For typical elliptical galaxies, Petrosian magnitudes can fall shy of
the total magnitude by a couple of tenths of a magnitude.  Application
of our method to the {\sl SDSS} Petrosian magnitudes yielded total
(S\'ersic) magnitudes accurate to a couple of hundredths of a
magnitude.  Moreover, it actually provided a more accurate measurement
of this quantity than obtained from direct S\'ersic fits to the data
(Blanton et al.\ 2005).

We have shown how the galaxy concentration can also be used to
determine the effective half--light radius $R_{\rm e}$ from the
Petrosian radius $R_{\rm P}$, and thereby additionally leading to a
determination of the associated surface brightness terms.  We
also provide magnitude corrections under the assumption that
galaxy light--profiles truncate at 5 effective radii.
 
In regard to the {\sl SDSS} data, Blanton et al.\ (2005) showed how
their S\'ersic fits to objects with increasingly higher values of $n$
resulted in increasingly underestimated effective radii.  For example,
their code obtained radii that were $\sim$80\% of the true effective
radii for simulated objects with $n=4.5$ ($R_{90}/R_{50}=3.4$).  Due
to this, the concentration--corrected effective radii derived from the
Petrosian 50\% radii are in fact preferable to those obtained from the
direct S\'ersic fits.  The same is obviously true for the associated
surface brightness terms.

In passing we warn that the use of $3R_{\rm P}$, with $1/\eta(R_{\rm
P})=0.5$, misses more than half of an $R^{1/4}$ profile's flux and
should therefore not be used in the analysis of bright elliptical
galaxies.  The use of $2R_{\rm P}$, with $1/\eta(R_{\rm P})=0.2$, as
justified by the {\sl SDSS} consortium, is suitably appropriate.  The
optimal choice of concentration remains an outstanding issue.

While the corrective scheme outlined here is general to any choice of
Petrosian index and aperture, the tabulated values we have provided
are specific to two commonly used Petrosian systems, namely that used
by CAS (Conselice 2003) and the setup adopted by the SDSS
collaboration (York et al.\ 2000).

\acknowledgments  
We are grateful for Michael Blanton's prompt and helpful replies to 
our assorted queries.  We also wish to thank the anonymous referee 
for their detailed comments. 
CJC acknowledges support from an NSF postdoctoral fellowship. 
MAB acknowledges support from NSF grant AST-0307417. 
SMC acknowledges support from STScI grant HST-AR-9917. 
SPD acknowledges financial support from the Australian
Research Council through DP0451426.

Funding for the Sloan Digital Sky Survey ({\sl SDSS}) has been
provided by the Alfred P. Sloan Foundation, the Participating
Institutions, the National Aeronautics and Space Administration, the
National Science Foundation, the U.S. Department of Energy, the
Japanese Monbukagakusho, and the Max Planck Society.
The {\sl SDSS} Web site is http://www.sdss.org/

\begin{deluxetable}{lcccccccc}
\tablewidth{0pt}
\tablecaption{Petrosian Radii, Concentration, and Magnitude Corrections\label{Table1}}
\tablehead
{
\colhead{$n$}  &  \colhead{$R_{\rm P}$}  &  \colhead{$R_{50}$}  &
\colhead{$R_{90}/R_{50}$}  &  \colhead{$R_{50}/R_{90}$}  &
\colhead{$\mu_{\rm e}-\mu_{50}$}  &  
\colhead{$\langle\mu\rangle_{\rm e}-\langle\mu\rangle_{50}$}  &
\colhead{$\Delta M_{\infty}$}  &  \colhead{$\Delta M_{5R_{\rm e}}$} \\
\colhead{}  &  \colhead{[$R_{\rm e}$]}  &  \colhead{[$R_{\rm e}$]}  & 
\colhead{}  & \colhead{}  &  
\colhead{mag arcsec$^{-2}$}  &  \colhead{mag arcsec$^{-2}$}  &
\colhead{mag}  &  \colhead{mag}  \\
\colhead{(1)}  &  \colhead{(2)}  &  \colhead{(3)}  & 
\colhead{(4)}  &  \colhead{(5)}  &  \colhead{(6)}  &
\colhead{(7)}  &  \colhead{(8)}  &  \colhead{(9)} 
}
\startdata
\multicolumn{9}{c}{Standard Petrosian: Aperture equals 
$2R_{\rm P}(1/\eta=0.2)$, $\eta$ given in Eq.~\ref{Eqeta}. } \\
0.1  &  1.57   & 1.00  &  1.40  &  0.71  & 0.00  & 0.00  &   0.00   &  0.00  \\
0.2  &  1.70   & 1.00  &  1.51  &  0.66  & 0.00  & 0.00  &   0.00   &  0.00  \\
0.3  &  1.80   & 1.00  &  1.62  &  0.62  & 0.00  & 0.00  &   0.00   &  0.00  \\
0.5  &  1.96   & 1.00  &  1.82  &  0.55  & 0.00  & 0.00  &   0.00   &  0.00  \\
0.7  &  2.06   & 1.00  &  2.02  &  0.49  & 0.00  & 0.00  &   0.00   &  0.00  \\
1.0  &  2.16   & 0.99  &  2.29  &  0.44  & 0.01  & 0.01  &   0.01   &  0.00  \\
2.0  &  2.20   & 0.93  &  2.90  &  0.34  & 0.13  & 0.09  &   0.06   &  0.02  \\
3.0  &  2.04   & 0.84  &  3.23  &  0.31  & 0.36  & 0.26  &   0.13   &  0.04  \\
4.0  &  1.82   & 0.73  &  3.42  &  0.29  & 0.63  & 0.48  &   0.20   &  0.07  \\
5.0  &  1.59   & 0.62  &  3.56  &  0.28  & 0.94  & 0.74  &   0.28   &  0.11  \\
6.0  &  1.36   & 0.53  &  3.65  &  0.27  & 1.27  & 1.03  &   0.35   &  0.15  \\
7.0  &  1.15   & 0.44  &  3.72  &  0.27  & 1.62  & 1.33  &   0.43   &  0.20  \\
8.0  &  0.97   & 0.37  &  3.78  &  0.26  & 1.98  & 1.65  &   0.50   &  0.25  \\
9.0  &  0.81   & 0.31  &  3.83  &  0.26  & 2.35  & 1.98  &   0.57   &  0.30  \\
10.0 &  0.67   & 0.25  &  3.87  &  0.26  & 2.73  & 2.33  &   0.64   &  0.36  \\
%
\tableline
\multicolumn{9}{c}{{\sl SDSS} Petrosian: Aperture equals
$2R_{\rm P}(1/\eta=0.2)$, $\eta$ given in Eq.~\ref{PetSDSS}.} \\
0.1  &  1.63   & 1.00  & 1.40   &  0.71  & 0.00  & 0.00  &   0.00   &  0.00  \\
0.2  &  1.72   & 1.00  & 1.51   &  0.66  & 0.00  & 0.00  &   0.00   &  0.00  \\
0.3  &  1.81   & 1.00  & 1.62   &  0.62  & 0.00  & 0.00  &   0.00   &  0.00  \\
0.5  &  1.94   & 1.00  & 1.82   &  0.55  & 0.00  & 0.00  &   0.00   &  0.00  \\
0.7  &  2.03   & 1.00  & 2.02   &  0.49  & 0.00  & 0.00  &   0.00   &  0.00  \\
1.0  &  2.11   & 0.99  & 2.29   &  0.44  & 0.01  & 0.01  &   0.01   &  0.01  \\
2.0  &  2.12   & 0.93  & 2.88   &  0.35  & 0.15  & 0.10  &   0.06   &  0.02  \\
3.0  &  1.94   & 0.82  & 3.17   &  0.32  & 0.38  & 0.28  &   0.14   &  0.05  \\
4.0  &  1.71   & 0.71  & 3.35   &  0.30  & 0.68  & 0.52  &   0.22   &  0.09  \\
5.0  &  1.48   & 0.61  & 3.47   &  0.29  & 1.00  & 0.79  &   0.30   &  0.13  \\
6.0  &  1.25   & 0.51  & 3.55   &  0.28  & 1.35  & 1.09  &   0.38   &  0.18  \\
7.0  &  1.05   & 0.42  & 3.62   &  0.28  & 1.72  & 1.41  &   0.46   &  0.23  \\
8.0  &  0.88   & 0.35  & 3.67   &  0.27  & 2.09  & 1.74  &   0.54   &  0.29  \\
9.0  &  0.72   & 0.29  & 3.71   &  0.27  & 2.48  & 2.09  &   0.61   &  0.34  \\
10.0 &  0.60   & 0.24  & 3.74   &  0.27  & 2.87  & 2.45  &   0.69   &  0.40  \\
%
\tableline
\multicolumn{9}{c}{CAS Petrosian: Aperture equals
$1.5R_{\rm P}(1/\eta=0.2)$, $\eta$ given in Eq.~\ref{Eqeta}.} \\
     &         &       & $R_{80}/R_{20}$ &  $R_{20}/R_{80}$  &   &  &   &   \\
0.1  &  1.57   & 1.00  & 2.04   &  0.49  & 0.00  & 0.00  &   0.00   &   0.00  \\
0.2  &  1.70   & 1.00  & 2.17   &  0.46  & 0.00  & 0.00  &   0.00   &   0.00  \\
0.3  &  1.80   & 1.00  & 2.33   &  0.43  & 0.00  & 0.00  &   0.00   &   0.00  \\
0.5  &  1.96   & 1.00  & 2.68   &  0.37  & 0.00  & 0.00  &   0.00   &   0.00  \\
0.7  &  2.06   & 0.99  & 3.03   &  0.33  & 0.01  & 0.01  &   0.01   &   0.01  \\
1.0  &  2.16   & 0.97  & 3.53   &  0.28  & 0.05  & 0.03  &   0.03   &   0.03  \\
2.0  &  2.20   & 0.88  & 4.96   &  0.20  & 0.26  & 0.17  &   0.12   &   0.08  \\
3.0  &  2.04   & 0.76  & 6.16   &  0.16  & 0.54  & 0.39  &   0.20   &   0.11  \\
4.0  &  1.82   & 0.65  & 7.20   &  0.14  & 0.86  & 0.66  &   0.29   &   0.16  \\
5.0  &  1.59   & 0.55  & 8.12   &  0.12  & 1.20  & 0.95  &   0.37   &   0.20  \\
6.0  &  1.36   & 0.46  & 8.96   &  0.11  & 1.56  & 1.26  &   0.45   &   0.25  \\
7.0  &  1.15   & 0.38  & 9.72   &  0.10  & 1.92  & 1.58  &   0.53   &   0.30  \\
8.0  &  0.97   & 0.31  & 10.43  &  0.10  & 2.30  & 1.92  &   0.60   &   0.35  \\
9.0  &  0.81   & 0.26  & 11.08  &  0.09  & 2.68  & 2.27  &   0.68   &   0.41  \\
10.0 &  0.67   & 0.21  & 11.67  &  0.09  & 3.07  & 2.62  &   0.75   &   0.46 
\enddata
\tablecomments{
Col.(1): S\'ersic index $n$. 
Col.(2): Petrosian radius $R_{\rm P}$ such that $1/\eta(R_{\rm P})=0.2$. 
Col.(3): Radius containing 50\% of the Petrosian flux. 
Col.(4): Concentration index defined by the ratio of radii that 
contain 90\% \& 50\% (and 80\& \& 20\%) of the Petrosian flux. 
Col.(5): The inverse of column (4).
Col.(6): Difference between the surface brightness at $R_{\rm e}$
and the value at $R_{50}$.
Col.(7): Difference between the mean surface brightness inside the radii 
$R_{\rm e}$ and $R_{50}$.
Col.(8): Difference between the Petrosian magnitude
and the total magnitude (obtained by integrating the S\'ersic
profile to infinity). 
Col.(9): Difference between the Petrosian magnitude
and the S\'ersic magnitude if the profile is truncated at $5R_{\rm e}$. 
}
\end{deluxetable}


\begin{references}
\reference{Abaz2}Abazajian, K., et al.\ 2004, ApJ, 625, 613
\reference{BJC00}Bershady, M.A., Jangren, A., \& Conselice, C.J.\ 2000, AJ, 119, 2645
\reference{BLK98}Bershady, M.A., Lowenthal, J.D., \& Koo, D.C.\ 1998, ApJ, 505, 50
\reference{BVFD5}Bland-Hawthorn, J., Vlaji\'c, M., Freeman, K.C., \& Drain, B.T.\ 2005, ApJ, in press (astro--ph/0503488)
\reference{Bet01}Blanton, M., et al.\ 2001, AJ, 121, 2358
\reference{Bet03}Blanton, M., et al.\ 2003a, ApJ, 592, 819
\reference{Bet03}Blanton, M., et al.\ 2003b, ApJ, 594, 186
\reference{Bet05}Blanton, M., et al.\ 2005, AJ, 129, 2562
\reference{CCD93}Caon, N., Capaccioli, M., \& D'Onofrio, M.\ 1993, MNRAS, 265, 1013
\reference{Cet05}Chang, R., et al.\ 2005, MNRAS, submitted (astro-ph/0502117)
\reference{Con02}Conselice, C.J., Gallagher III, J.S., \& Wyse, R.F.G.\ 2002, AJ, 123, 2246
\reference{Cet03}Conselice, C.J., et al.\ 2003, AJ, 125, 66
\reference{Con03}Conselice, C.J.\ 2003, ApJS, 147, 1
\reference{Cio91}Ciotti, L.\ 1991, A\&A, 249, 99
\reference{Cet01}Cross, N.J.G., et al.\ 2001, MNRAS, 324, 825
\reference{Cet04}Cross, N.J.G., Driver, S.P., Liske, J., Lemon, D.J., Peacock, J.A., Cole, S., Norberg, P., \& Sutherland, W.J.\ 2004, MNRAS, 349, 576
\reference{Dal98}Dalcanton, J.J.\ 1998, ApJ, 495, 251
\reference{dGd04}de Souza, R.E., Gadotti, D.A., \& dos Anjos, S.\ 2004, ApJS, 153, 411
\reference{DeR05}De Rijcke, S., Michielsen, D., Dejonghe, H., Zeilinger, W.W., \& Hau, G.K.T.\ 2005, A\&A, in press (astro--ph/0412553) 
\reference{DaD87}Djorgovski, S., \& Davis, M.\ 1987, ApJ, 313, 59
\reference{DaS81}Djorgovski, S., \& Spinrad, H.\ 1981, ApJ, 251, 417
\reference{Dre87}Dressler, A., et al.\ 1987, ApJ, 313, 42
\reference{Det05}Driver, S.P., Liske J., Cross, N.J.G., De Propris R., \& Allen P.D.\ 2005, MNRAS, in press (astro--ph/0503228)
\reference{Dut05}Dutton, A.A., van den Bosch, F.C., Courteau, S., \& Dekel, A.\ 2005, (astro--ph/0501256)
\reference{EBP05}Erwin, P., Beckman, J.E., \& Pohlen, M.\ 2005, ApJL, in press (astro--ph/0505216)
\reference{FaJ76}Faber, S.M., \& Jackson, R.E.\ ApJ, 204, 668
\reference{Goto3}Goto, T., et al.\ 2003, PASJ, 55, 739
\reference{Gra02}Graham, A.W.\ 2002, MNRAS, 334, 721
\reference{GaD05}Graham, A.W., \& Driver, S.\ 2005, PASA, in press
\reference{Get01}Graham, A.W., Erwin, P., Caon, N., \& Trujillo, I.\ 2001, ApJ, 563, L11
\reference{GaG03}Graham, A.W., \& Guzm\'an, R.\ 2003, AJ, 125, 2936
\reference{GTC01}Graham, A.W., Trujillo, I., \& Caon, N., 2001, AJ, 122, 1707
\reference{GaO75}Gunn, J.E., \& Oke, J.B.\ 1975, ApJ, 195, 255
\reference{Ket03}Kauffmann, G., et al.\ 2003, MNRAS, 341, 54
\reference{Kor77}Kormendy, J.\ 1977, ApJ, 218, 333
\reference{LaS01}Lubin, L.M., \& Sandage, A.\ 2001, AJ, 122, 1084
\reference{MaG05}Matkovi\'c, A., \& Guzm\'an, R.\ 2005, ApJ, submitted
\reference{McI05}McIntosh, D.I., et al.\ 2005, ApJ, submitted (astro--ph/0411772)
\reference{MaD04}McLure, R.J., \& Dunlop, J.S.\ 2004, MNRAS, 352, 1390
\reference{Nak03}Nakamura O., Fukugita M., Yasuda N., Loveday J., Brinkmann J., Schneider D. P., Shimasaku K., SubbaRao M.\ 2003, AJ, 125, 1682
\reference{NaJ03}Narayan, C.A., \& Jog, C.J.\ 2003, A\&A, 407, L59
\reference{Net02}Norberg, P., et al.\ 2002, MNRAS, 336, 907
\reference{PHIHW}Peng, C.Y., Ho, L.C., Impey, C.D., \& Hans-Walter, W.\ 2002, AJ, 124, 266
\reference{Pet76}Petrosian, V.\ 1976, ApJ, 209, L1
\reference{Poh04}Pohlen M., Beckman, J.E., H\"uttemeister, S., Knapen, J.H., Erwin, P., \& Dettmar, R.-J.\ 2004, in Penetrating Bars Through Masks of Cosmic Dust, eds., D.L.\ Block, I.\ Puerari, K.C.\ Freeman, R.\ Groess, E.K.\ Block, A\&SS, 319, 713
\reference{PaH97}Prugniel, Ph., \& H\'eraudeau, Ph.\ 1997, A\&ASS, 128, 299
\reference{SaP90}Sandage, A., \& Perelmuter, J.-M.\ ApJ, 1990, 350, 481
\reference{Sco01}Scodeggio, M.\ 2001, AJ, 121, 2413
\reference{Ser63} S\'ersic, J.L.\ 1963, Boletin de la Asociacion Argentina de Astronomia, vol.6, p.41
\reference{Ser68}S\'ersic, J.L.\ 1968, Atlas de galaxias australes
\reference{Set04}Shankar, F., Salucci, P., Granato, G.L., De Zotti, G., \& Danese, L.\ 2004, MNRAS, 354, 1020
\reference{She03}Shen, S., et al.\ 2003, MNRAS, 343, 978
\reference{Shi01}Shimasaku, K., et al.\ 2001, AJ, 122, 1238
\reference{GIM2D}Simard, L., et al.\ 2002, ApJS, 142, 1
\reference{Str02}Strauss, M.A., et al.\ 2002, AJ, 124, 1810
\reference{Set01}Strateva, I., et al.\ 2001, AJ, 122, 1861
\reference{Tak99}Takamiya, M.\ 1999, ApJS, 122, 109
\reference{Tr01a}Trujillo, I., Graham, A.W., \& Caon, N.\ 2001a, MNRAS, 326, 869
\reference{Tr01b}Trujillo, I., Aguerri, J.A.L., Cepa, J., Guti\'errez, C.M.\ 2001b, MNRAS, 321, 269
\reference{Tr01c}Trujillo, I., Aguerri, J.A.L., Cepa, J., Guti\'errez, C.M.\ 2001c, MNRAS, 328, 977
\reference{TaF77}Tully, R.B., \& Fisher, J.R.\ 1977, A\&A, 54, 661
\reference{vdK01}van der Kruit, P.C.\ 2001, in Galaxy Disks and Disk Galaxies, G.\ Jos\'e, S.J.\ Funes, E.M.\ Corsini, ASP Conf.\ Ser., 230, 119
\reference{VSC00}Volonteri, M., Saracco, P., Chincarini, G.\ 2000, A\&AS, 145, 111
\reference{WKK94}Wirth, G.D., Koo, D.C., \& Kron, R.G.\ 1994, ApJ, 435, L105
\reference{Yet02}Yagi, M., Kashikawa, N., Sekiguchi, M., Doi, M., Yasuda, N., Shimasaku, K., \& Okamura, S.\ 2002, AJ, 123, 66
\reference{Yet01}Yasuda, N., et al.\ 2001, AJ, 122, 1104
\reference{SDSS0}York, D.G., et al.\ 2000, AJ, 120, 1579
\end{references}
\end{document}